\documentclass[conference]{IEEEtran}
\IEEEoverridecommandlockouts

\usepackage{cite}
\usepackage{amsmath,amssymb,amsfonts}
\usepackage{algorithmic}
\usepackage{graphicx}
\usepackage{textcomp}
\usepackage{xcolor}
\def\BibTeX{{\rm B\kern-.05em{\sc i\kern-.025em b}\kern-.08em
    T\kern-.1667em\lower.7ex\hbox{E}\kern-.125emX}}

\usepackage{ntheorem}
\theoremstyle{plain}
\theoremseparator{.}
\theoremheaderfont{\bfseries\upshape}
\theorembodyfont{\itshape}

\newtheorem{theorem}{Theorem}
\newtheorem{lemma}{Lemma}

\newtheorem{definition}{Definition}
\newtheorem{conjecture}{Conjecture}

\usepackage{comment}

\begin{document}

\title{Improved constructions of secondary structure avoidance codes for DNA sequences}

\author{
\textbf{Hui Chu}, \textbf{Chen Wang}, and \textbf{Yiwei Zhang} \\
\IEEEauthorblockA{Key Laboratory of Cryptologic Technology and Information Security of Ministry of Education, \\ School of Cyber Science and Technology, Shandong University, Qingdao, Shandong, 266237, China}
{chuhui@mail.sdu.edu.cn ~ cwang2021@mail.sdu.edu.cn ~ ywzhang@sdu.edu.cn}
\thanks{Research supported in part by National Key Research and Development Program
of China under Grant Nos. 2020YFA0712100, 2022YFA1004900, and 2021YFA1001000, in part by National Natural Science Foundation of China under Grant Nos. 12001323 and 12231014,
and in part by Shandong Provincial Natural Science Foundation under Grant No. ZR2021YQ46. Corresponding author: Yiwei Zhang.}
}

\maketitle

\begin{abstract}
In a DNA sequence, we have the celebrated Watson-Crick complement $\overline{T}=A$, $\overline{A}=T$, $\overline{C}=G$, and $\overline{G}=C$.
Given an integer $m\ge 2$, a secondary structure in a DNA sequence refers to the existence of two non-overlapping reverse complement consecutive subsequences of length $m$,
denoted as $\boldsymbol{x}=(x_1, \dots, x_m)$ and $\boldsymbol{y}=(y_1, \dots, y_m)$, such that $x_i=\overline{y_{m-i+1}}$ for $1\leq i \leq m$.
The property of secondary structure avoidance (SSA) forbids a sequence to contain such reverse complement subsequences,
and it is a key criterion in the design of single-stranded DNA sequences for DNA computing and storage.
In this paper, we improve on a recent result of Nguyen et al., by introducing explicit constructions of secondary structure avoidance codes and analyzing the capacity for any given $m$. In particular, our constructions have optimal rate 1.1679bits/nt and 1.5515bits/nt when $m=2$ and $m=3$, respectively.
\end{abstract}

\section{Introduction}\label{Sec:intro}

DNA sequences have become a promising medium for both computation and storage  \cite{A94Sci,CGK12Sci,GBCDLSB13Nat}.
A DNA sequence is a quaternary sequence with the alphabet set $\{A,T,C,G\}$, which represents the four nucleotides:
Adenine ($A$), Thymine ($T$), Cytosine ($C$), and Guanine ($G$).
The celebrated {\it Watson-Crick complement} indicates that $\overline{T}=A$, $\overline{A}=T$, $\overline{C}=G$, and $\overline{G}=C$.

\begin{figure}[!h]
  \centering
  \includegraphics[width=.4\textwidth]{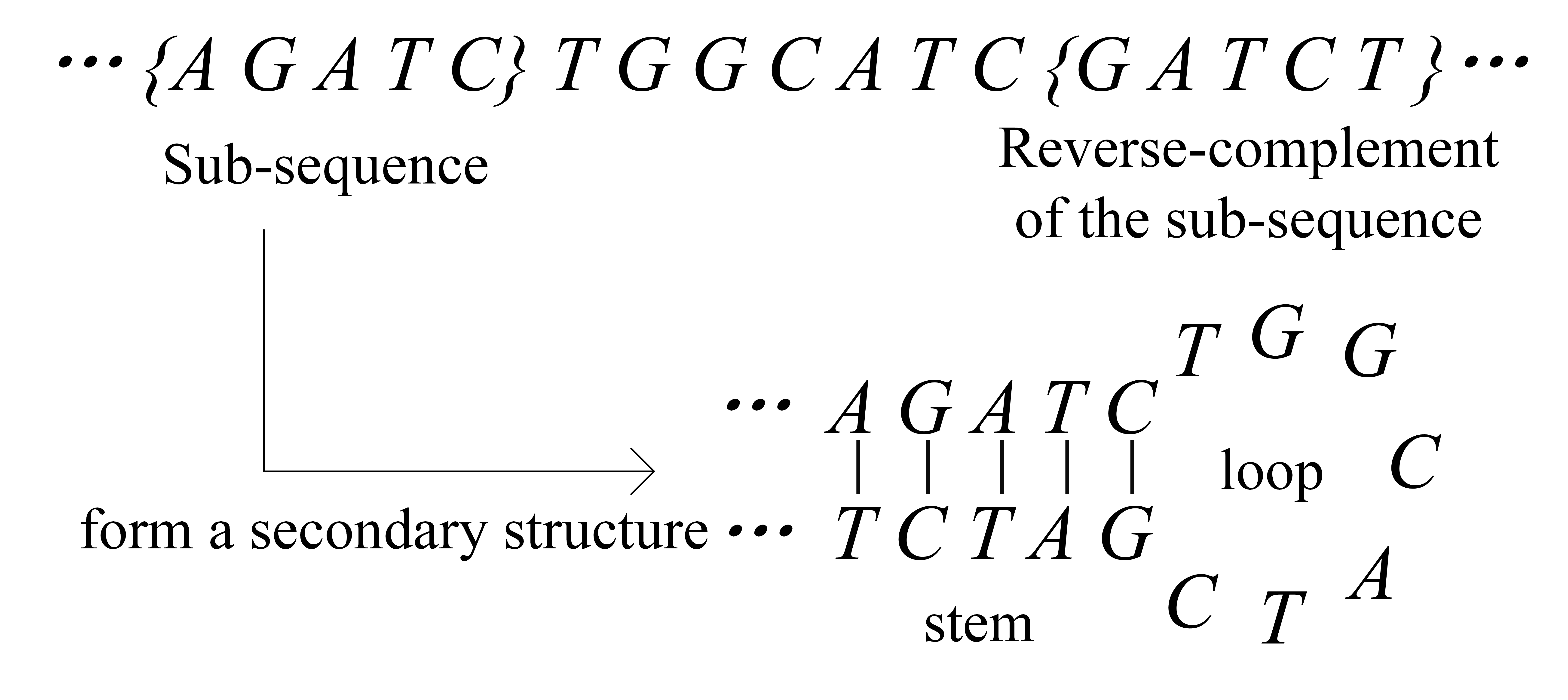}\\
  \caption{A stem-loop secondary structure}\label{Fig:emp}
\end{figure}

In a single-stranded DNA sequence, a secondary structure refers to the existence of two non-overlapping reverse complement consecutive subsequences,
which will make the sequence fold back onto itself by complementary base pair hybridization and form a stem-loop structure (see Figure \ref{Fig:emp}).
The existence of stem-loop structures makes the sequence less active.
Thus, in both DNA computation and DNA storage, it is desirable to use sequences avoiding secondary structure \cite{MK06Cod,TKGM18TIT,CMW19ISIT}.
While there have been some dynamic programming techniques \cite{BFM86Pro,NJ80Pro} to efficiently check whether a given DNA sequence possesses secondary structure,
and there have also been some works \cite{MK06Cod,MK05ISIT} discussing the sequence design criteria for reducing the possibility of secondary structure appearances,
it would be more natural to have a coding theoretic solution by directly constructing as many secondary structure avoidance (SSA) sequences as possible.

Given the sequence length $n$ and the stem length $m\geq 2$, an $m$-SSA sequence does not contain two non-overlapping reverse complement consecutive subsequences of length $m$.
An $m$-SSA code is a set of $m$-SSA sequences.
It should be remarked that in current applications of DNA computation and storage,
biochemists usually consider the secondary structure problem with stem length $m\geq 6$ and loop length at least 4 \cite{BGI}.
In our mathematical formulation, we consider a more general setting by allowing $m\geq 2$ and ignoring the parameter of the loop length
(essentially it does not affect the rate of the code).
The SSA sequences have been studied in \cite{MK06Cod,BB21ICL,MK05ISIT,BB23NCC,NCKDI23ArXiv}. In particular, the recent paper \cite{NCKDI23ArXiv} applies a symbol-composition constrained technique to construct $3$-SSA codes with rate 1.3031bits/nt,
which improves on the prior result 1.1609bits/nt constructed by a block concatenation method from \cite{BB21ICL}.

In this paper, we further analyze the capacity and improve the constructions of SSA codes. 
In particular, our constructions and capacity analysis together lead to optimal rates 1.1679bits/nt and 1.5515bits/nt for the cases $m=2$ and $m=3$, respectively.
The rest of the paper is organized as follows.
In Section \ref{Sec:pre} we introduce the mathematical notations.
In Section \ref{Sec:odd} we give a construction when $m$ is odd and in particular we can explicitly calculate the rate of our construction for $m\in\{3,5\}$.
In Section \ref{Sec:upb} we provide a method to determine the the capacity of SSA codes,
which requires a brute-force search over a so-called generating set $S$ for an $m$-SSA code.
In Section \ref{Sec:constr} we further discuss the strategy to choose a good generating set $S$, and compute the rate of the code by a standard approach in constrained coding theory. Conclusions and discussions are given in Section \ref{Sec:concl}.

\section{Preliminary}\label{Sec:pre}

Let $\mathcal{D}=\{A,T,C,G\}$ be the quaternary alphabet set for DNA sequences,
where the Watson-Crick complement indicates that $\overline{T}=A$, $\overline{A}=T$, $\overline{C}=G$, and $\overline{G}=C$.
We use bold letters to represent the sequences and plain letters to represent the symbols in the sequences, such as $\boldsymbol{x}=(x_1,x_2,\dots,x_n)$.
In this paper, whenever we talk about subsequences we mean consecutive subsequences.
Consider two non-overlapping subsequences $\boldsymbol{y}=\boldsymbol{x}[i;m]\triangleq(x_i,x_{i+1},\dots,x_{i+m-1})$ and $\boldsymbol{z}=\boldsymbol{x}[j;m]=(x_j,x_{j+1},\dots,x_{j+m-1})$ of some $\boldsymbol{x}$, where $1\leq i < i+m-1 < j \leq n-m+1$.
$\boldsymbol{y}$ and $\boldsymbol{z}$ are called reverse-complement, denoted as $\boldsymbol{y}=\mathrm{RC}(\boldsymbol{z})$, if $x_{i+t}=\overline{x_{j+m-1-t}}$ for all $0\leq t\leq m-1$.

\begin{definition}
Given integers $2<m\leq n$, a DNA sequence $\boldsymbol{x}\in\mathcal{D}^n$ is said to be an {\it $m$-secondary structure avoidance sequence} (abbreviated as an $m$-SSA sequence),
if it does not contain any two non-overlapping reverse-complement subsequences of length $m$\footnote{In \cite{NCKDI23ArXiv} it is defined as ``of length at least $m$", but it is straightforward to show that the two definitions are equivalent.}.
A code $\mathcal{C}\subseteq \mathcal{D}^n$ is called an $m$-SSA code if every codeword in $\mathcal{C}$ is an $m$-SSA sequence.
\end{definition}

The {\it code rate} of $\mathcal{C}$ is defined as $\frac{1}{n}\log(|\mathcal{C}|)$ with logarithm base 2. It measures the number of bits encoded in each DNA nucleotide. For example, $\mathcal{C}=\mathcal{D}^n$ has the largest rate 2bits/nt.

\begin{definition}
Given an integer $m\geq 2$, let $A(n;m)$ be the largest size of an $m$-SSA code of length $n$. Define the capacity of $m$-SSA codes as
$$c_m=\lim_{n\rightarrow \infty} \frac{1}{n} \log (A(n;m)).$$
\end{definition}

Next we present two known constructions of $3$-SSA codes from \cite{BB21ICL} and \cite{NCKDI23ArXiv}.

\begin{theorem}\cite{BB21ICL}
Set $S=\{AA,AC,CA,CC,TC\}$. Let $\mathcal{C}$ be a code of even length such that each codeword is formed by a block concatenation using blocks from $S$. Then $\mathcal{C}$ is a $3$-SSA code of size $5^{n/2}$. Thus, $c_3\geq \frac{1}{2}\log 5=1.1609$ bits/nt.
\end{theorem}

\begin{theorem}\cite{NCKDI23ArXiv}
Let $\mathcal{C}$ be the set of all DNA sequences such that any subsequence of length $3$ contains at least one $A$ and does not contain any $T$. Let $f(n)$ be the size of such a code. It holds that $f(1)=3$, $f(2)=9$, $f(3)=19$, and the linear recurrence relation is $f(n)=f(n-1)+2f(n-2)+4f(n-3)$. Then $\mathcal{C}$ is a $3$-SSA code of size $\rho^n$, where $\rho=2.4675$ is the largest real root of the corresponding characteristic equation $x^3-x^2-2x-4=0$. Thus, $c_3\geq \log (\rho)=1.3031$ bits/nt.
\end{theorem}

\section{A construction when $m$ is odd}\label{Sec:odd}

In this section we propose a simple construction of $m$-SSA codes when $m$ is odd.

\begin{definition}
A DNA sequence $\boldsymbol{x}$ is called {\it $TC$-$m$-dominant}, if for every subsequence of length $m$, the sum of appearances of $T$ and $C$ is larger than $m/2$.
\end{definition}

\begin{theorem}
When $m$ is odd, a $TC$-$m$-dominant sequence must be an $m$-SSA sequence.
Thus a code $\mathcal{C}$ consisting of all $TC$-$m$-dominant sequences of length $n$ is an $m$-SSA code.
\end{theorem}

\begin{IEEEproof}
Consider any two non-overlapping subsequences $\boldsymbol{y}$ and $\boldsymbol{z}$ in a $TC$-$m$-dominant sequence $\boldsymbol{x}$.
In $\boldsymbol{z}$, the sum of appearances of $T$ and $C$ is larger than $m/2$.
Note that the complements of $T$ and $C$ are $A$ and $G$, respectively.
Thus, if $\boldsymbol{y}=\mathrm{RC}(\boldsymbol{z})$, then sum of appearances of $A$ and $G$ should be larger than $m/2$ in $\boldsymbol{y}$,
which contradicts to the assumption that $\boldsymbol{x}$ is a $TC$-$m$-dominant sequence.
\end{IEEEproof}

It should be remarked that we can define $AC$-$m$-dominant, $TG$-$m$-dominant, and $AG$-$m$-dominant sequences analogously, and all of them can be shown to be $m$-SSA sequences. The reason we choose to use $TC$-$m$-dominant sequences as representatives is a biochemical observation that $T$ and $C$ are relatively cheaper and more efficient in the current DNA synthesis process \cite{BGI}. Moreover, while all the four families of sequences could constitute an $m$-SSA code, for computing the rate it suffices to only consider one of them.

Next, in order to compute the rate of our construction, we map the quaternary sequences to binary sequences by mapping $T$ and $C$ into $1$, and mapping $A$ and $G$ into $0$. It is routine to check this is a $2^n$-to-one mapping. The set of $TC$-$3$-dominant DNA sequences are mapped into binary sequences such that every length-$3$ subsequence has weight at least $2$, and we refer to them as {\it good sequences}. Now, let $f_3(n)$ be the number of good sequences, and we can derive the following linear recurrence relation.

\begin{theorem}
Let $f_3(n)$ be the number of good sequences, which are binary sequences such that every length-$3$ subsequence has weight at least $2$. For $n\geq 4$ it holds that
$$f_3(n)=f_3(n-1)+f_3(n-3).$$
\end{theorem}

\begin{IEEEproof}
On one hand, if a good sequence of length $n$ starts with $1$, then the $1$ could be followed by any good sequence of length $n-1$. This accounts for the $f_3(n-1)$ part.
On the other hand, if a good sequence of length $n$ starts with $0$, then it must start with $011$, and then any good sequence of length $n-3$ could follow. This accounts for the $f_3(n-3)$ part.
\end{IEEEproof}

Therefore, the number of good sequences is approximately $f_3(n)\approx \rho^n$, where $\rho=1.4656$ is the largest real root of the corresponding characteristic equation $x^3-x^2-1=0$. Since each good sequence corresponds to $2^n$ $TC$-$m$-dominant DNA sequences, we have $A(n;3)\geq \rho^n\cdot 2^n$ and thus $c_3\geq 1+\log (\rho)=1.5514$ bits/nt.

Similarly, we switch to the case $m=5$, and map the $TC$-$5$-dominant DNA sequences into binary sequences such that every length-$5$ subsequence has weight at least $3$. Again by abuse of notation we refer to them as good sequences.

\begin{theorem}
Let $f_5(n)$ be the number of good sequences, which are binary sequences such that every length-$5$ subsequence has weight at least $3$. For $n\geq 11$ it holds that
\begin{small}
$$f_5(n)=f_5(n-1)+f_5(n-3)+2f_5(n-5)-f_5(n-8)-f_5(n-10).$$
\end{small}
\end{theorem}

\begin{IEEEproof}
Again we classify good sequences according to their prefix.
\begin{itemize}
  \item Good sequences starting with $1$: $f_5(n-1)$.
  \item Good sequences starting with $00$ must start with $00111$, and then any good sequence of length $n-5$ could follow. The number of this class is then $f_5(n-5)$.
  \item Good sequences starting with $010$ must start with $01011$, and then any good sequence of length $n-5$ could follow, except for those starting with $00$. Moreover, a good sequence of length $n-5$ starting with $00$ must start with $00111$, and is then followed by any good sequence of length $n-10$. The number of this class is then $f_5(n-5)-f_5(n-10)$.
  \item Good sequences starting with $011$, could be followed by any good sequence of length $n-3$ except for those starting with $00$. Moreover, a good sequence of length $n-3$ starting with $00$ must start with $00111$, and is then followed by any good sequence of length $n-8$. The number of this class is then $f_5(n-3)-f_5(n-8)$.
\end{itemize}
In sum we get the desired recurrence relation.
\end{IEEEproof}

Therefore, the number of good sequences is approximately $f_5(n)\approx \rho^n$, where $\rho=1.6222$ is the largest real root of the corresponding characteristic equation $x^{10}-x^9-x^7-2x^5+x^2+1=0$. Since each good sequence corresponds to $2^n$ $TC$-$5$-dominant DNA sequences, we have $A(n;5)\geq \rho^n\cdot 2^n$ and thus $c_3\geq 1+\log (\rho)=1.6980$ bits/nt.

Theoretically the procedure could continue. However, finding the recurrence relation even for $m=7$ is already a tedious and nontrivial task. We do not bother to go on this computation. Instead, we can turn to the standard spectral radius method in constrained coding theory by computer search (see Section \ref{Sec:constr}).

\section{Upper bound analysis of the capacity}\label{Sec:upb}

In \cite{NCKDI23ArXiv}, a trivial upper bound is given as $c_m\leq \frac{1}{m}\log(\frac{4^m}{2})$, which leads to $c_2\leq 1.5$ and $c_3\leq 1.67$.
In this section, we analyze the upper bound of the capacity for $m$-SSA codes by a standard approach in constrained coding theory,
i.e., we use a state transition digraph to characterize the constrained system. We start with the following definition.

\begin{definition}
Given $m\geq 2$, let $\mathcal{S}_m=\{S_1,S_2,\dots,S_{\tau(m)}\}$ be a family of sets, where each set $S_i\subseteq \mathcal{D}^m$, $1\leq i \leq \tau(m)$, is a maximal set of sequences such that in $S_i$ there are no two reverse complement sequences $\boldsymbol{x}$ and $\boldsymbol{y}$ (not necessarily distinct).
\end{definition}

For example, when $m=2$, each $S_i$ cannot contain the self reverse complement sequences $\{TA,AT,CG,GC\}$, and contains exactly one out of each reverse complement pairs: $\{TT,AA\}$, $\{TC, GA\}$, $\{TG, CA\}$, $\{AC,GT\}$, $\{AG,CT\}$, and $\{CC,GG\}$. Thus $\mathcal{S}_2$ is a family of $\tau(2)=2^6$ sets.

For any $\boldsymbol{x}\in \mathcal{D}^n$, let $M(\boldsymbol{x})=\{\boldsymbol{x}[i;m]:1\leq i\leq n-m+1\}$ be the multiset of all length-$m$ subsequences of $\boldsymbol{x}$. Consider the two codes defined as follows.

\begin{definition}
For any $S\in\mathcal{S}_m$,
Let $C_n(S)\subseteq \mathcal{D}^n$ be the set of all sequences $\boldsymbol{x}$ such that every sequence in $M(\boldsymbol{x})$ belongs to $S$.
Let $\widetilde{C}_n(S)\subseteq \mathcal{D}^n$ be the set of all sequences $\boldsymbol{x}$ such that each subsequence in $M(\boldsymbol{x})\setminus S$ has multiplicity at most $2m-1$ in $M(\boldsymbol{x})$.
We use
$C_n(\mathcal{S}_m)$ and $\widetilde{C}_n(\mathcal{S}_m)$ to denote $\bigcup_{S\in\mathcal{S}_m}C_n(S)$ and $\bigcup_{S\in\mathcal{S}_m}\widetilde{C}_n(S)$, respectively.
\end{definition}

\begin{lemma}\label{Lem:sandwich}
Let $\mathcal{C}$ be the optimal $m$-SSA code. Then it holds that
$C_n(\mathcal{S}_m)\subseteq \mathcal{C} \subseteq \widetilde{C}_n(\mathcal{S}_m)$.
\end{lemma}
\begin{IEEEproof}
By definition, it is routine to check every sequence in $C_n(S)$ does not contain reverse complement subsequences of length $m$, for any $S\in\mathcal{S}_m$. Therefore $C_n(\mathcal{S}_m)\subseteq \mathcal{C}$.

For any $\boldsymbol{x}\in \mathcal{C}$, let $M'(\boldsymbol{x})$ be the set of subsequences in $M(\boldsymbol{x})$ appearing at least $2m$ times. We claim that $M'(\boldsymbol{x})$ is a subset of at least one $S\in\mathcal{S}_m$. Suppose otherwise, then in $M'(\boldsymbol{x})$ there exists two reverse complement sequences $\boldsymbol{y}$ and $\boldsymbol{z}$. Note that there are at most $2m-1$ sequences have intersecting coordinates with $\boldsymbol{z}$ (calculating with multiplicity and including $\boldsymbol{z}$ itself), and $\boldsymbol{y}$ appears at least $2m$ times as a subsequence of $\boldsymbol{x}$. Thus, we can find some $\boldsymbol{y}$ which does not have overlap with $\boldsymbol{z}$, which makes $\boldsymbol{x}$ not an $m$-SSA sequence. Therefore, every sequence in $M(\boldsymbol{x})\setminus S$ appears at most $2m-1$ times, so $\boldsymbol{x}$ belongs to $\widetilde{C}_n(S)\subseteq \widetilde{C}_n(\mathcal{S}_m)$.
\end{IEEEproof}

Next we show that for any $S\in \mathcal{S}_m$, the capacity of $C_n(S)$ and $\widetilde{C}_n(S)$ are in fact equal.

\begin{lemma}\label{Lem:tilde}
For any $S\in\mathcal{S}_m$, if
$$\lim_{n\rightarrow \infty} \frac{1}{n}\log(|C_n(S)|)=\lambda,$$
then it holds that
$$\limsup_{n\rightarrow \infty} \frac{1}{n}\log(|\widetilde{C}_n(S)|)=\lambda.$$
\end{lemma}
\begin{IEEEproof}
(Outline) We calculate the number of sequences in $\widetilde{C}_n(S)$ as follows.
Such a sequence contains at most a constant number of subsequences not in $S$, and each such subsequence appears at most a constant number of times.
Therefore, out of the $n$ coordinates we can first select the positions of the sequences not belonging to $S$ (the number of possible ways is still a constant number only dependent on $m$), and then the remaining blank positions are divided into several intervals of length $n_1,n_2,n_3,\dots$
Each of the interval should be filled with sequences in $C_{n_i}(S)$ for some length $n_i$, and the possibility for such an interval is bounded by $O(\lambda^{n_i})$ when $n_i$ is sufficiently large. Thus, $|\widetilde{C}_n(S)|\leq O_m(1) \times \Pi_{i\geq 1}\lambda^{n_1} = O_m(1)\times \lambda^n$, where $O_m(1)$ represents a constant only dependent on $m$.
\end{IEEEproof}

Note that when $m$ is a fixed integer and $n$ approaches infinity, $\tau(m)$ is a constant number at most $4^m/2$. Thus, the capacity of $C_n(\mathcal{S}_m)$ is determined by the largest value among the capacities of $\{C_n(S):S\in\mathcal{S}_m\}$. The same holds for $\widetilde{C}_n(\mathcal{S}_m)$. Thus Lemma \ref{Lem:sandwich} and Lemma \ref{Lem:tilde} together lead to the following result.

\begin{theorem}
The capacities of $m$-SSR codes, $C_n(\mathcal{S}_m)$, and $\widetilde{C}_n(\mathcal{S}_m)$ are in fact equal, and this value could be computed by searching for the largest value among $\{C_n(S):S\in\mathcal{S}_m\}$.
\end{theorem}

To sum up, we prove that the capacity of $m$-SSR codes can be derived by searching for the largest capacity $\{C_n(S):S\in\mathcal{S}_m\}$, where $C_n(S)$ is the set of length $n$ sequences such that every subsequence of length $m$ belongs to $S$. The capacity of $C_n(S)$ can be computed by a standard constrained coding approach, to be introduced in the next section.

\section{Constructions of $m$-SSA codes}\label{Sec:constr}

We call $S$ a {\it generating set} for $C_n(S)$ and in this section we will discuss how to choose a good generating set $S$ which leads to $m$-SSA codes with good rate. We start with the case $m=2$. As has been explained in the previous section, $S$ contains exactly one out of the following six pairs: $\{TT,AA\}$, $\{TC, GA\}$, $\{TG, CA\}$, $\{AC,GT\}$, $\{AG,CT\}$, and $\{CC,GG\}$.

Say we choose $S=\{TT,TC,TG,GT,CT,CC\}$. $C_n(S)$ is the set of length $n$-sequences such that any subsequence of length 2 belongs to $S$. The size, or the capacity, of $C_n(S)$, can be determined by a standard constrained coding theory approach as follows. Build a directed graph with $6$ vertices labelled by $S$, and there is a directed arc from one vertex to the other (not necessarily distinct), if the second coordinate of the former equals the first coordinate of the latter. The adjacency matrix of this directed graph is as follows (rows and columns indexed by the order $TT,TC,TG,GT,CT,CC$):
\begin{center}
$\left(
  \begin{array}{cccccc}
    1 & 1 & 1 & 0 & 0 & 0 \\
    0 & 0 & 0 & 0 & 1 & 1 \\
    0 & 0 & 0 & 1 & 0 & 0 \\
    1 & 1 & 1 & 0 & 0 & 0 \\
    1 & 1 & 1 & 0 & 0 & 0 \\
    0 & 0 & 0 & 0 & 1 & 1 \\
  \end{array}
\right)$
\end{center}
Then the capacity of $C_n(S)$ equals $\log(\rho)=1.1679$ bits/nt, where $\rho=2.247$ is the spectral radius of the matrix above. Since there are only 64 different choices of $S$ for $m=2$ (in fact, with some additional work one can show that there are only 8 equivalent classes), we can use a brute-force algorithm to calculate the spectral radius for every possible $S$. It turns out $1.1679$ bits/nt is indeed optimal.

For a general $m$, after picking the generating set $S$, the directed graph is formed by drawing an arc from $(x_1,\dots,x_m)$ to $(y_1,\dots,y_m)$ whenever $x_{i+1}=y_i$ for all $1\leq i\leq m-1$, and the spectral method can be applied. Thus, what remains is to find the best $S$ for each $m$, and it seems a nontrivial task. When $m$ is odd, essentially our construction in Section \ref{Sec:odd} is to pick $S$ as the set of all length-$m$ sequences which are $TC$-dominant (the sum of appearances of $T$ and $C$ is over $m/2$). For $m=3$, we use a brute-force search for all possible generating sets, and verify that 1.5514bits/nt is indeed optimal. For $m\in\{7,9,11\}$, our construction has rate 1.7698, 1.8131, 1.8423 bits/nt accordingly. We pose the following conjecture.

\begin{conjecture}
When $m$ is odd, the generating set $S$ with all $TC$-dominant length-$m$ sequences results in $m$-SSA codes with the largest rate.
\end{conjecture}

When $m$ is even, the choice of $S$ gets complicated since we have many length-$m$ sequences with exactly one half of the symbols to be $T$ and $C$.
We discuss the cases $m=4$ and $m=6$ in the following two subsections, respectively.

\subsection{Discussions on $m=4$}

In this subsection we discuss how to choose the generating set $S$ for $m=4$.

Among the $4^4=256$ sequences of length $4$, $16$ of them (e.g. $ACGT$) are self reverse complement, and the rest are divided into $120$ reverse complement pairs. A brute-force search requires checking over $2^{120}$ possible choices of $S$ and is thus impractical. Moreover, taking equivalence into consideration does not have much help.

To choose the generating set $S$, we first pick those $TC$-dominant sequences (the sum of appearances of $T$ and $C$ is 4 or 3). For example, in a reverse complement pair $\{TCCA,TGGA\}$ we choose $TCCA$ and ignore $TGGA$.

We still have to decide our choices for the sequences containing exactly a sum of two appearances of $T$ and $C$.
The choice is made by further analyzing the positions of $T$ and $C$.
Now denote $T$ and $C$ as $1$, and denote $A$ and $G$ as $0$.

Our strategy goes on with the following steps:

\begin{itemize}
  \item Between a reverse complement pair, where one is of the form $1001$ and the other one is of the form $0110$, we give our priority to $0110$. The reason is that such vertices will have more arcs with the already chosen $TC$-dominant vertices (which are of the form $1111$, $1110$, $1101$, $1011$, $0111$). Observe that a vertex in the form $0110$ is for sure to have an incoming arc from a vertex of the form $1011$, and an outgoing arc to a vertex of the form $1101$, and potential connections with $0011$ and $1100$, if they exist. However, for a vertex in the form $1001$, they only have potential connections with $0011$ and $1100$, but no more arcs since vertices in the form of $0100$ and $0010$ do not exist. Thus, we choose the vertices in the form $0110$ and neglect $1001$.
  \item Between a reverse complement pair, where one is of the form $0011$ and the other one is of the form $1100$, it does not matter what we choose. Observe that vertices of the form $1100$ do not have outgoing arcs, because $1000$ and $1001$ are already banned in our choice. Similarly, the vertices of the form $0011$ do not have incoming edges since $0001$ and $1001$ are banned. While allowing these vertices in the generating set $S$ will certainly result in more $m$-SSA sequences, they could be neglected when computing the capacity and the spectral radius of the adjacency matrix.
  \item Finally there are $12$ pairs left. In $6$ pairs both sequences are of the form $1010$ (e.g. $TACA$ and $TGTA$), and in the other $6$ pairs both sequences are of the form $0101$ (e.g. $GCAT$ and $ATGC$). Among these $12$ pairs we perform a brute-force search and find the desired choice as $\{CACA, TACA, CGCA, CATA, TACG, CACG, \\ACAC, GCAC, ATAC, ACGC, GCAT, ACAT\}$. Here a reasonable explanation for the choice of these 12 sequences is that they have the largest number of induced arcs within them.
 \end{itemize}

Following the previous strategy, the generating set $S$ results in a rate $\log(3.0190)=1.5940$ bits/nt and we conjecture it to be optimal (it's impractical to perform a brute-force over all the $2^{120}$ choices and 1.5940 is indeed the best result we have found by computer search so far).

\subsection{Discussions on $m=6$}

In this subsection we discuss how to choose the generating set $S$ for $m=6$. Following the same idea as in the previous subsection,
we first pick all the $TC$-$6$-dominating sequences. For the sequences containing exactly a sum of three appearances of $T$ and $C$, our strategy goes on with the following steps:

\begin{itemize}
  \item Between a reverse complement pair, where one is of the form $100011$ and the other one if of the form $001110$, we give our priority to $001110$ since it has more connections with the $TC$-$6$-dominating sequences. Similarly, we choose sequences of the forms $010110$, $011010$, and $011100$, whereas neglect sequences of the forms $100101$, $101001$, and $110001$, accordingly.
  \item Sequences of the forms $000111$, $111000$, $001011$, $110100$ are self reverse complement. They can all be neglected since they have no incoming arcs or no outgoing arcs, and thus do not contribute to the rate of the code.
  \item  Between a reverse complement pair, where one is of the form $001101$ and the other one if of the form $010011$, we choose $001101$ because sequences of the form $010011$ do not have incoming arcs. Similarly, we choose sequences of the form $101100$ and neglect sequences of the form $110010$.
\end{itemize}

We pause here to mention that, at this step the generating set already results in a spectral radius $3.2443$ and thus a code rate $\log(3.2443)=1.6979$ bits/nt, which already beats the result $1.5312$ bits/nt derived by \cite{NCKDI23ArXiv}. We can still proceed a little bit more:

\begin{itemize}
  \item Consider the sequences of the forms $010101$ and $101010$. First delete the self reverse complement ones such as $TACGTA$ (altogether there are 16 such sequences). We still have $56$ reverse complement pairs left in this class and it is still impractical to run a brute-force search. It is natural to guess that we should pick these sequences such that they have the most number of induced arcs within them.
  \item Consider the sequences of the forms $011001$ and $100110$. First delete the self reverse complement ones such as $TGGCCA$ (altogether there are 16 such sequences). We still have $56$ reverse complement pairs left in this class.
\end{itemize}

We haven't come up with a good strategy for the sequences in these two classes. So far our best result by brute-force search has rate $\log(3.3097)=1.7267$ bits/nt.

\section{Conclusion}\label{Sec:concl}

In this paper we analyze the constructions and the capacity of $m$-SSR codes, which are beneficial for the applications in DNA computing and DNA storage. 
The rate of our construction is summarized in Table \ref{Tab:summary}, and in particular we have derived optimal rates when $m\in \{2,3\}$ with the aid of computer search.
We conjecture that $TC$-$m$-dominant sequences give rise to the codes with optimal rate when $m$ is odd, but so far there is a lack of theoretical analysis.
Furthermore, we still need to analyze the strategy for choosing the generating set when $m$ is even. 

\begin{table}[h]
\begin{center}
\scalebox{0.9}{
  \begin{tabular}{|c|c|c|c|c|c|c|c|}
    \hline
    $m$ & 2 & 3 & 4 & 5 & 7 & 9 & 11 \\\hline
    Rate & 1.1679 & 1.5515 & 1.5940 & 1.6980 & 1.7698 & 1.8131 & 1.8423 \\
    \hline
  \end{tabular}
}  
\end{center}
\caption{Rate of our construction for SSA codes (bits/nt)}\label{Tab:summary}
\end{table}

\section*{Acknowledgements}
We deeply thank Dr. Xianger Jiang and Mr. Haoling Zhang from BGI-research, for valuable discussions on DNA chemical synthesis and introducing the biochemical backgrounds of this problem.

\end{document}